\documentclass[twocolumn,amsmath,amssymb,10pt,superscriptaddress,a4paper,letterpaper,fleqn]{revtex4-1}
\usepackage{amssymb}
\usepackage{epsfig}
\usepackage{graphicx}
\usepackage{dcolumn}
\usepackage{array}
\usepackage{bm}
\usepackage{fancyheadings}
\usepackage{longtable}
\usepackage{multirow}
\usepackage{float}
\usepackage{afterpage}  
\usepackage{color}

\setlongtables
\usepackage[breaklinks=true,linkbordercolor={1 1 1}]{hyperref}

\begin{document}
\newlength{\wdo}
\newcommand{\stroke}[1]{{$#1$}%
\settowidth{\wdo}{${#1}$} {\kern-\wdo}%
\partialvartstrokedint}

\newenvironment{ibox}[1]%
{\vskip 1.0em
\framebox[\columnwidth][r]{%
\begin{minipage}[c]{\columnwidth}%
\vspace{-1.0em}%
#1%
\end{minipage}}}
{\vskip 1.0em}

\newcommand{\iboxed}[1]{%
\vskip 1.0em
\framebox[\columnwidth][r]{%
\begin{minipage}[c]{\columnwidth}%
\vspace{-1.0em}
#1%
\end{minipage}}
\vskip 1.0em}

\newcommand{\fitbox}[2]{%
\vskip 1.0em
\begin{flushright}
\framebox[{#1}][r]{%
\begin{minipage}[c]{\columnwidth}%
\vspace{-1.0em}
#2%
\end{minipage}}
\end{flushright}
\vskip 1.0em}

\newcommand{\iboxeds}[1]{%
\vskip 1.0em
\begin{equation}
\fbox{%
\begin{minipage}[c]{1mm}%
\vspace{-1.0em}
#1%
\end{minipage}}
\end{equation}
\vskip 1.0em}

\def\Xint#1{\mathchoice
   {\XXint\displaystyle\textstyle{#1}}%
   {\XXint\textstyle\scriptstyle{#1}}%
   {\XXint\scriptstyle\scriptscriptstyle{#1}}%
   {\XXint\scriptscriptstyle\scriptscriptstyle{#1}}%
   \!\int}
\def\XXint#1#2#3{{\setbox0=\hbox{$#1{#2#3}{\int}$}
     \vcenter{\hbox{$#2#3$}}\kern-.5\wd0}}
\def\ddashint{\Xint=}
\def\dashint{\Xint-}

\def\mathbi#1{\textbf{\em #1}}

\newcommand{\alps}{\ensuremath{\alpha_s}}
\newcommand{\qbar}{\bar{q}}
\newcommand{\ubar}{\bar{u}}
\newcommand{\dbar}{\bar{d}}
\newcommand{\sbar}{\bar{s}}
\newcommand{\beq}{\begin{equation}}
\newcommand{\eeq}{\end{equation}}
\newcommand{\beqa}{\begin{eqnarray}}
\newcommand{\eeqa}{\end{eqnarray}}
\newcommand{\gs}{g_{\pi NN}}
\newcommand{\gw}{f_\pi}
\newcommand{\mq}{m_Q}
\newcommand{\mn}{m_N}
\newcommand{\mpi}{m_\pi}
\newcommand{\mrho}{m_\rho}
\newcommand{\momg}{m_\omega}
\newcommand{\bb}{\langle}
\newcommand{\kb}{\rangle}
\newcommand{\xvec}{\mathbf{x}}
\newcommand{\st}{\ensuremath{\sqrt{\sigma}}}
\newcommand{\Bvec}{\mathbf{B}}
\newcommand{\rvec}{\mathbf{r}}
\newcommand{\Rvec}{\mathbf{R}}
\newcommand{\kvec}{\mathbf{k}}
\newcommand{\pvec}{\mathbf{p}}
\newcommand{\Pvec}{\mathbf{P}}
\newcommand{\vvec}{\mathbf{v}}
\newcommand{\Vvec}{\mathbf{V}}
\newcommand{\bvec}[1]{\ensuremath{\mathbf{#1}}}
\newcommand{\bra}[1]{\ensuremath{\bb#1|}}
\newcommand{\ket}[1]{\ensuremath{|#1\kb}}
\newcommand{\pbra}[1]{\ensuremath{(#1|}}
\newcommand{\pket}[1]{\ensuremath{|#1)}}
\newcommand{\gft}{\ensuremath{\gamma_{FT}}}
\newcommand{\gfv}{\ensuremath{\gamma_5}}
\newcommand{\bfalp}{\ensuremath{\bm{\alpha}}}
\newcommand{\bfbeta}{\ensuremath{\bm{\beta}}}
\newcommand{\bfeps}{\ensuremath{\bm{\epsilon}}}
\newcommand{\lag}{{\lambda_\gamma}}
\newcommand{\lao}{{\lambda_\omega}}
\newcommand{\lN}{\lambda_N}
\newcommand{\lM}{\lambda_M}
\newcommand{\lB}{\lambda_B}
\newcommand{\epslag}{\ensuremath{\bm{\epsilon}_{\lag}}}
\newcommand{\bfept}{\ensuremath{\tilde{\bm{\epsilon}}}}
\newcommand{\bfgam}{\ensuremath{\bm{\gamma}}}
\newcommand{\bfnab}{\ensuremath{\bm{\nabla}}}
\newcommand{\bflambda}{\ensuremath{\bm{\lambda}}}
\newcommand{\bfmu}{\ensuremath{\bm{\mu}}}
\newcommand{\bfphi}{\ensuremath{\bm{\phi}}}
\newcommand{\bfvphi}{\ensuremath{\bm{\varphi}}}
\newcommand{\bfpi}{\ensuremath{\bm{\pi}}}
\newcommand{\bfsig}{\ensuremath{\bm{\sigma}}}
\newcommand{\bftau}{\ensuremath{\bm{\tau}}}
\newcommand{\bfpsi}{\ensuremath{\bm{\psi}}}
\newcommand{\bfdelta}{\ensuremath{\bm{\delta}}}
\newcommand{\bfrho}{\ensuremath{\bm{\rho}}}
\newcommand{\bfth}{\ensuremath{\bm{\theta}}}
\newcommand{\bfchi}{\ensuremath{\bm{\chi}}}
\newcommand{\bfxi}{\ensuremath{\bm{\xi}}}
\newcommand{\bfR}{\ensuremath{\bvec{R}}}
\newcommand{\bfP}{\ensuremath{\bvec{P}}}
\newcommand{\bfJ}{{\mathbi{J}}}
\newcommand{\bfj}{{\mathbi{j}}}
\newcommand{\Rcm}{\ensuremath{\bvec{R}_{CM}}}
\newcommand{\spup}{\uparrow}
\newcommand{\spd}{\downarrow}
\newcommand{\up}{\uparrow}
\newcommand{\dn}{\downarrow}
\newcommand{\hbarom}{\frac{\hbar^2}{m_Q}}
\newcommand{\half}{\ensuremath{\frac{1}{2}}}
\newcommand{\thalf}{\ensuremath{\frac{3}{2}}}
\newcommand{\fhalf}{\ensuremath{\frac{5}{2}}}
\newcommand{\shalf}{\ensuremath{{\tfrac{1}{2}}}}
\newcommand{\sqtr}{\ensuremath{{\tfrac{1}{4}}}}
\newcommand{\sphalf}{\ensuremath{\genfrac{}{}{0pt}{1}{+}{}\!\tfrac{1}{2}}}
\newcommand{\smhalf}{\ensuremath{\genfrac{}{}{0pt}{1}{-}{}\!\tfrac{1}{2}}}
\newcommand{\sthalf}{\ensuremath{{\tfrac{3}{2}}}}
\newcommand{\spthalf}{\ensuremath{{\tfrac{+3}{2}}}}
\newcommand{\smthalf}{\ensuremath{{\tfrac{-3}{2}}}}
\newcommand{\sfhalf}{\ensuremath{\tfrac{5}{2}}}
\newcommand{\sshalf}{\ensuremath{\tfrac{7}{2}}}
\newcommand{\third}{\ensuremath{\frac{1}{3}}}
\newcommand{\tthird}{\ensuremath{\frac{2}{3}}}
\newcommand{\sthird}{\ensuremath{\tfrac{1}{3}}}
\newcommand{\stthird}{\ensuremath{\tfrac{2}{3}}}
\newcommand{\vnn}{\ensuremath{\hat{v}_{NN}}}
\newcommand{\vij}{\ensuremath{\hat{v}_{ij}}}
\newcommand{\vik}{\ensuremath{\hat{v}_{ik}}}
\newcommand{\argonne}{\ensuremath{v_{18}}}
\newcommand{\lqcd}{\ensuremath{\mathcal{L}_{QCD}}}
\newcommand{\lqed}{\ensuremath{\mathscr{L}_{QED}}}
\newcommand{\lgf}{\ensuremath{\mathcal{L}_g}}
\newcommand{\lqm}{\ensuremath{\mathcal{L}_q}}
\newcommand{\lqg}{\ensuremath{\mathcal{L}_{qg}}}
\newcommand{\nn}{\ensuremath{N\!N}}
\newcommand{\nnn}{\ensuremath{N\!N\!N}}
\newcommand{\qq}{\ensuremath{qq}}
\newcommand{\qqq}{\ensuremath{qqq}}
\newcommand{\qqb}{\ensuremath{q\bar{q}}}
\newcommand{\hpnd}{\ensuremath{H_{\pi N\Delta}}}
\newcommand{\hpqq}{\ensuremath{H_{\pi qq}}}
\newcommand{\hpqqa}{\ensuremath{H^{(a)}_{\pi qq}}}
\newcommand{\hpqqe}{\ensuremath{H^{(e)}_{\pi qq}}}
\newcommand{\hint}{\ensuremath{H_{\rm int}}}
\newcommand{\fpnn}{\ensuremath{f_{\pi\! N\!N}}}
\newcommand{\fenn}{\ensuremath{f_{\eta\! N\!N}}}
\newcommand{\gsnn}{\ensuremath{g_{\sigma\! N\!N}}}
\newcommand{\gpnn}{\ensuremath{g_{\pi\! N\!N}}}
\newcommand{\fpnd}{\ensuremath{f_{\pi\! N\!\Delta}}}
\newcommand{\grpg}{\ensuremath{g_{\rho\pi\gamma}}}
\newcommand{\gopg}{\ensuremath{g_{\omega\pi\gamma}}}
\newcommand{\fmqq}{\ensuremath{f_{M\! qq}}}
\newcommand{\gmqq}{\ensuremath{g_{M\! qq}}}
\newcommand{\fpqq}{\ensuremath{f_{\pi qq}}}
\newcommand{\gpqq}{\ensuremath{g_{\pi qq}}}
\newcommand{\feqq}{\ensuremath{f_{\eta qq}}}
\newcommand{\gonn}{\ensuremath{g_{\omega N\!N}}}
\newcommand{\gonna}{\ensuremath{g^t_{\omega N\!N}}}
\newcommand{\grnn}{\ensuremath{g_{\rho N\!N}}}
\newcommand{\gopr}{\ensuremath{g_{\omega\pi\rho}}}
\newcommand{\grnp}{\ensuremath{g_{\rho N\!\pi}}}
\newcommand{\grpp}{\ensuremath{g_{\rho\pi\pi}}}
\newcommand{\Lpnn}{\ensuremath{\Lambda_{\pi\! N\! N}}}
\newcommand{\Lonn}{\ensuremath{\Lambda_{\omega N\! N}}}
\newcommand{\Lonna}{\ensuremath{\Lambda^t_{\omega N\! N}}}
\newcommand{\Lrnn}{\ensuremath{\Lambda_{\rho N\! N}}}
\newcommand{\Lopr}{\ensuremath{\Lambda_{\omega\pi\rho}}}
\newcommand{\Lrpp}{\ensuremath{\Lambda_{\rho\pi\pi}}}
\newcommand{\getaqq}{\ensuremath{g_{\eta qq}}}
\newcommand{\fsqq}{\ensuremath{f_{\sigma qq}}}
\newcommand{\gsqq}{\ensuremath{g_{\sigma qq}}}
\newcommand{\piqq}{\ensuremath{{\pi\! qq}}}
\newcommand{\ylm}{\ensuremath{Y_\ell^m}}
\newcommand{\ylmc}{\ensuremath{Y_\ell^{m*}}}
\newcommand{\ebh}[1]{\hat{\bvec{e}}_{#1}}
\newcommand{\kbh}{\hat{\bvec{k}}}
\newcommand{\nbh}{\hat{\bvec{n}}}
\newcommand{\pvbh}{\hat{\bvec{p}}}
\newcommand{\qbh}{\hat{\bvec{q}}}
\newcommand{\Xbh}{\hat{\bvec{X}}}
\newcommand{\rbh}{\hat{\bvec{r}}}
\newcommand{\xbh}{\hat{\bvec{x}}}
\newcommand{\ybh}{\hat{\bvec{y}}}
\newcommand{\zbh}{\hat{\bvec{z}}}
\newcommand{\betabh}{\hat{\bfbeta}}
\newcommand{\tbh}{\hat{\bfth}}
\newcommand{\pbh}{\hat{\bfvphi}}
\newcommand{\dt}{\Delta\tau}
\newcommand{\kmag}{|\bvec{k}|}
\newcommand{\pmag}{|\bvec{p}|}
\newcommand{\qmag}{|\bvec{q}|}
\newcommand{\oas}{\ensuremath{\mathcal{O}(\alpha_s)}}
\newcommand{\vtxb}{\ensuremath{\Lambda_\mu(p',p)}}
\newcommand{\vtxp}{\ensuremath{\Lambda^\mu(p',p)}}
\newcommand{\pwqp}{e^{i\bvec{q}\cdot\bvec{r}}}
\newcommand{\pwqm}{e^{-i\bvec{q}\cdot\bvec{r}}}
\newcommand{\gsa}[1]{\ensuremath{\bb#1\kb_0}}
\newcommand{\oer}[1]{\mathcal{O}\left(\frac{1}{\qmag^{#1}}\right)}
\newcommand{\nub}[1]{\overline{\nu^{#1}}}
\newcommand{\epf}{E_\bvec{p}}
\newcommand{\epfp}{E_{\bvec{p}'}}
\newcommand{\eka}{E_{\alpha\kappa}}
\newcommand{\ekaq}{(E_{\alpha\kappa})^2}
\newcommand{\ekap}{E_{\alpha'\kappa}}
\newcommand{\ekpa}{E+{\alpha\kappa_+}}
\newcommand{\ekma}{E_{\alpha\kappa_-}}
\newcommand{\ekp}{E_{\kappa_+}}
\newcommand{\ekm}{E_{\kappa_-}}
\newcommand{\ekpap}{E_{\alpha'\kappa_+}}
\newcommand{\ekmap}{E_{\alpha'\kappa_-}}
\newcommand{\yjm}[1]{\mathcal{Y}_{jm}^{#1}}
\newcommand{\ysa}[3]{\mathcal{Y}_{#1,#2}^{#3}}
\newcommand{\yjsl}[2]{\mathcal{Y}_{#1}^{#2}}
\newcommand{\yss}[2]{\mathcal{Y}_{#1}^{#2}}
\newcommand{\Dj}{\ensuremath{\mathscr{D}}}
\newcommand{\ysc}{\tilde{y}}
\newcommand{\enm}{\varepsilon_{NM}}
\newcommand{\Scg}[6]
	{\ensuremath{S^{#1}_{#4}\:\vphantom{S}^{#2}_{#5}
 	 \:\vphantom{S}^{#3}_{#6}\,}}
\newcommand{\Kmat}[6]
	{\ensuremath{K\left[\begin{array}{ccc} 
	#1 & #2 & #3 \\ #4 & #5 & #6\end{array}\right]}}
\newcommand{\irt}{\ensuremath{\frac{1}{\sqrt{2}}}}
\newcommand{\sirt}{\ensuremath{\tfrac{1}{\sqrt{2}}}}
\newcommand{\irth}{\ensuremath{\frac{1}{\sqrt{3}}}}
\newcommand{\sirth}{\ensuremath{\tfrac{1}{\sqrt{3}}}}
\newcommand{\irs}{\ensuremath{\frac{1}{\sqrt{6}}}}
\newcommand{\sirs}{\ensuremath{\tfrac{1}{\sqrt{6}}}}
\newcommand{\tors}{\ensuremath{\frac{2}{\sqrt{6}}}}
\newcommand{\stors}{\ensuremath{\tfrac{2}{\sqrt{6}}}}
\newcommand{\rtoth}{\ensuremath{\sqrt{\frac{2}{3}}}}
\newcommand{\rthot}{\ensuremath{\frac{\sqrt{3}}{2}}}
\newcommand{\ithrt}{\ensuremath{\frac{1}{3\sqrt{2}}}}
\newcommand{\Tg}{\ensuremath{\mathsf{T}}}
\newcommand{\irrep}[1]{\ensuremath{\mathbf{#1}}}
\newcommand{\cirrep}[1]{\ensuremath{\overline{\mathbf{#1}}}}
\newcommand{\Fij}{\ensuremath{\hat{F}_{ij}}}
\newcommand{\Fqij}{\ensuremath{\hat{F}^{(qq)}_{ij}}}
\newcommand{\Fsij}{\ensuremath{\hat{F}^{(qs)}_{ij}}}
\newcommand{\Opij}{\mathcal{O}^p_{ij}}
\newcommand{\fpij}{f_p(r_{ij})}
\newcommand{\titj}{\bftau_i\cdot\bftau_j}
\newcommand{\sisj}{\bfsig_i\cdot\bfsig_j}
\newcommand{\Sij}{S_{ij}}
\newcommand{\LS}{\bvec{L}_{ij}\cdot\bvec{S}_{ij}}
\newcommand{\TT}{\Tg_i\cdot\Tg_j}
\newcommand{\chet}{\ensuremath{\chi ET}}
\newcommand{\chpt}{\ensuremath{\chi PT}}
\newcommand{\chsy}{\ensuremath{\chi\mbox{symm}}}
\newcommand{\lchi}{\ensuremath{\Lambda_\chi}}
\newcommand{\lcon}{\ensuremath{\Lambda_{QCD}}}
\newcommand{\dcpsi}{\ensuremath{\bar{\psi}}}
\newcommand{\dcbfpsi}{\ensuremath{\bar{\bfpsi}}}
\newcommand{\dc}[1]{\ensuremath{\overline{#1}}}
\newcommand{\dcpsip}{\ensuremath{\bar{\psi}^{(+)}}}
\newcommand{\psip}{\ensuremath{{\psi}^{(+)}}}
\newcommand{\dcpsim}{\ensuremath{\bar{\psi}^{(-)}}}
\newcommand{\psim}{\ensuremath{{\psi}^{(-)}}}
\newcommand{\llo}{\ensuremath{\mathcal{L}^{(0)}_{\chet}}}
\newcommand{\lchet}{\ensuremath{\mathcal{L}_{\chi}}}
\newcommand{\hchet}{\ensuremath{\mathcal{H}_{\chi}}}
\newcommand{\Hd}{\ensuremath{\mathcal{H}}}
\newcommand{\Dmu}{\ensuremath{\mathcal{D}_\mu}}
\newcommand{\Dsl}{\ensuremath{\slashed{\mathcal{D}}}}
\newcommand{\comm}[2]{\ensuremath{\left[#1,#2\right]}}
\newcommand{\acomm}[2]{\ensuremath{\left\{#1,#2\right\}}}
\newcommand{\ev}[1]{\ensuremath{\bb\hat{#1}\kb}}
\newcommand{\exv}[1]{\ensuremath{\bb{#1}\kb}}
\newcommand{\evt}[1]{\ensuremath{\bb{#1}(\tau)\kb}}
\newcommand{\evm}[1]{\ensuremath{\bb{#1}\kb_M}}
\newcommand{\evv}[1]{\ensuremath{\bb{#1}\kb_V}}
\newcommand{\ovl}[2]{\ensuremath{\bb{#1}|{#2}\kb}}
\newcommand{\pd}{\partial}
\newcommand{\pnpd}[2]{\frac{\partial{#1}}{\partial{#2}}}
\newcommand{\pppd}[1]{\frac{\partial{\hphantom{#1}}}{\partial{#1}}}
\newcommand{\fnfd}[2]{\frac{\delta{#1}}{\delta{#2}}}
\newcommand{\rfnfd}[2]{\frac{\bfdelta{#1}}{\bfdelta{#2}}}
\newcommand{\fdfd}[1]{\frac{\delta}{\delta{#1}}}
\newcommand{\rfdfd}[1]{\frac{\overleftarrow{\delta}}{\delta{#1}}}
\newcommand{\brfdfd}[1]{\frac{{\bfdelta}}{\bfdelta{#1}}}
\newcommand{\plmu}{\partial_\mu}
\newcommand{\plnu}{\partial_\nu}
\newcommand{\pumu}{\partial^\mu}
\newcommand{\punu}{\partial^\nu}
\newcommand{\mcdf}{\delta^{(4)}(p_f-p_i-q)}
\newcommand{\ecdf}{\delta(E_f-E_i-\nu)}
\newcommand{\tr}{\mbox{Tr }}
\newcommand{\lxr}{\ensuremath{SU(2)_L\times SU(2)_R}}
\newcommand{\gV}[2]{\ensuremath{(\gamma^{-1})^{#1}_{\hphantom{#1}{#2}}}}
\newcommand{\gVd}[2]{\ensuremath{\gamma^{#1}_{\hphantom{#1}{#2}}}}
\newcommand{\LpV}[1]{\ensuremath{\Lambda^{#1}V}}
\newcommand{\hatH}{\ensuremath{\hat{H}}}
\newcommand{\hath}{\ensuremath{\hat{h}}}
\newcommand{\eht}{\ensuremath{e^{-\tau\hat{H}}}}
\newcommand{\ehdt}{\ensuremath{e^{-\Delta\tau\hat{H}}}}
\newcommand{\ehtm}{\ensuremath{e^{-\tau(\hat{H}-E_V)}}}
\newcommand{\ehdtm}{\ensuremath{e^{-\Delta\tau(\hat{H}-E_V)}}}
\newcommand{\Oop}{\ensuremath{\mathcal{O}}}
\newcommand{\Sop}{\ensuremath{\mathcal{S}}}
\newcommand{\Jop}{\ensuremath{\mathcal{J}}}
\newcommand{\Gop}{\ensuremath{\hat{\mathcal{G}}}}
\newcommand{\SU}[1]{\ensuremath{SU({#1})}}
\newcommand{\U}[1]{\ensuremath{U({#1})}}
\newcommand{\proj}[1]{\ensuremath{\ket{#1}\bra{#1}}}
\newcommand{\su}[1]{\ensuremath{\mathfrak{su}({#1})}}
\newcommand{\ip}[2]{\ensuremath{\bvec{#1}\cdot\bvec{#2}}}
\newcommand{\norm}[1]{\ensuremath{\left| #1\right|^2}}
\newcommand{\rnorm}[1]{\ensuremath{\lvert #1\rvert}}
\newcommand{\pid}{\left(\begin{array}{cc} 1 & 0 \\ 0 & 1\end{array}\right)}
\newcommand{\psx}{\left(\begin{array}{cc} 0 & 1 \\ 1 & 0\end{array}\right)}
\newcommand{\psy}{\left(\begin{array}{cc} 0 & -i \\ i & 0\end{array}\right)}
\newcommand{\psz}{\left(\begin{array}{cc} 1 & 0 \\ 0 & -1\end{array}\right)}
\newcommand{\ua}{\uparrow}
\newcommand{\da}{\downarrow}
\newcommand{\deln}{\delta_{i_1 i_2\ldots i_n}}
\newcommand{\GabRR}{G_{\alpha\beta}(\bfR,\bfR')}
\newcommand{\GRR}{G(\bfR,\bfR')}
\newcommand{\GfRR}{G_0(\bfR,\bfR')}
\newcommand{\GRiR}{G(\bfR_i,\bfR_{i-1})}
\newcommand{\GRRs}[2]{G(\bfR_{#1},\bfR_{#2})}
\newcommand{\Gdgn}{\Gamma_{\Delta,\gamma N}}
\newcommand{\Gdgnb}{\overline\Gamma_{\Delta,\gamma N}}
\newcommand{\GJT}{\Gamma_{LS}^{JT}(k)}
\newcommand{\GJTa}[2]{\Gamma^{#1}_{#2}}
\newcommand{\GtwJTa}[2]{\tilde{\Gamma}_{#1}^{#2}}
\newcommand{\Gtw}{\tilde{\Gamma}}
\newcommand{\Gbar}{\overline{\Gamma}}
\newcommand{\Gtil}{\tilde{\Gamma}}
\newcommand{\Gpndb}{\overline{\Gamma}_{\pi N,\Delta}}
\newcommand{\GbNgn}{{\overline{\Gamma}}_{N^*,\gamma N}}
\newcommand{\GNgn}{\Gamma_{N^*,\gamma N}}
\newcommand{\GbNmb}{{\overline{\Gamma}}_{N^*,MB}}
\newcommand{\Lg}[2]{\ensuremath{L^{#1}_{\hphantom{#1}{#2}}}}
\newcommand{\psik}{\ensuremath{\left(\begin{matrix}\psi_1 \\ \psi_2\end{matrix}\right)}}
\newcommand{\psib}{\ensuremath{\left(\begin{matrix}\psi^*_1&\psi^*_2\end{matrix}\right)}}
\newcommand{\Gf}{\ensuremath{\frac{1}{E-H_0}}}
\newcommand{\Gv}{\ensuremath{\frac{1}{E-H_0-\vnres}}}
\newcommand{\Gx}{\ensuremath{\frac{1}{E-H_0-V}}}
\newcommand{\Gex}{\ensuremath{\mathcal{G}}}
\newcommand{\Gfpm}{\ensuremath{\frac{1}{E-H_0\pm i\epsilon}}}
\newcommand{\vres}{v_R}
\newcommand{\vnres}{v}
\newcommand{\tpz}{\ensuremath{^3P_0}}
\newcommand{\tres}{t_R}
\newcommand{\tsr}{t^R}
\newcommand{\tsnr}{t^{NR}}
\newcommand{\trest}{\tilde{t}_R}
\newcommand{\tnres}{t}
\newcommand{\Pt}{P_{12}}
\newcommand{\Sz}{\ket{S_0}}
\newcommand{\Sa}{\ket{S^{(-1)}_1}}
\newcommand{\Sb}{\ket{S^{(0)}_1}}
\newcommand{\Sc}{\ket{S^{(+1)}_1}}
\newcommand{\sbasis}{\ket{s_1 s_2; m_1 m_2}}
\newcommand{\Sbasis}{\ket{s_1 s_2; S M}}
\newcommand{\sket}[2]{\ket{{#1}\,{#2}}}
\newcommand{\sbra}[2]{\bra{{#1}\,{#2}}}
\newcommand{\psmket}{\ket{\bvec{p};s\,m}}
\newcommand{\cket}{\ket{\bvec{p};s_1 s_2\,m_1 m_2}}
\newcommand{\hket}{\ket{\bvec{p};s_1 s_2\,\lambda_1\lambda_2}}
\newcommand{\hkets}{\ket{s\,\lambda}}
\newcommand{\phkets}{\ket{\bvec{p};s\,\lambda}}
\newcommand{\klsjm}{\ket{p;\ell s; j m}}
\newcommand{\pq}{\bvec{p}_q}
\newcommand{\pqb}{\bvec{p}_{\qbar}}
\newcommand{\mps}[1]{\frac{d^3{#1}}{(2\pi)^{3/2}}}
\newcommand{\mpsf}[1]{\frac{d^3{#1}}{(2\pi)^{3}}}
\newcommand{\du}[1]{u_{\bvec{#1},s}}
\newcommand{\dv}[1]{v_{\bvec{#1},s}}
\newcommand{\cdu}[1]{\overline{u}_{\bvec{#1},s}}
\newcommand{\cdv}[1]{\overline{v}_{\bvec{#1},s}}
\newcommand{\dus}[2]{u_{\bvec{#1},{#2}}}
\newcommand{\dvs}[2]{v_{\bvec{#1},{#2}}}
\newcommand{\cdus}[2]{\overline{u}_{\bvec{#1},{#2}}}
\newcommand{\cdvs}[2]{\overline{v}_{\bvec{#1},{#2}}}
\newcommand{\bop}[1]{b_{\bvec{#1},s}}
\newcommand{\dop}[1]{d_{\bvec{#1},s}}
\newcommand{\bops}[2]{b_{\bvec{#1},{#2}}}
\newcommand{\dops}[2]{d_{\bvec{#1},{#2}}}
\newcommand{\mev}{\mbox{ MeV}}
\newcommand{\gev}{\mbox{ GeV}}
\newcommand{\fmi}{\mbox{ fm}}
\newcommand{\M}{\mathcal{M}}
\newcommand{\Smat}{\mathcal{S}}
\newcommand{\JLSTh}{JLST\lambda}
\newcommand{\Tpg}{T_{\pi N,\gamma N}}
\newcommand{\tpg}{t_{\pi N,\gamma N}}
\newcommand{\vmbmb}{\ensuremath{v_{M'B',MB}}}
\newcommand{\tmbgn}{\ensuremath{t_{MB,\gamma N}}}
\newcommand{\Tonon}{\ensuremath{T_{\omega N,\omega N}}}
\newcommand{\tonon}{\ensuremath{t_{\omega N,\omega N}}}
\newcommand{\tronon}{\ensuremath{t^R_{\omega N,\omega N}}}
\newcommand{\Tpnpn}{\ensuremath{T_{\pi N,\pi N}}}
\newcommand{\Tpnppn}{\ensuremath{T_{\pi\pi N,\pi N}}}
\newcommand{\Tonpn}{\ensuremath{T_{\omega N,\pi N}}}
\newcommand{\tonpn}{\ensuremath{t_{\omega N,\pi N}}}
\newcommand{\tronpn}{\ensuremath{t^R_{\omega N,\pi N}}}
\newcommand{\Tongn}{\ensuremath{T_{\omega N,\gamma N}}}
\newcommand{\tongn}{\ensuremath{t_{\omega N,\gamma N}}}
\newcommand{\trongn}{\ensuremath{t^R_{\omega N,\gamma N}}}
\newcommand{\vmbgn}{\ensuremath{v_{MB,\gamma N}}}
\newcommand{\vpngn}{\ensuremath{v_{\pi N,\gamma N}}}
\newcommand{\vongn}{\ensuremath{v_{\omega N,\gamma N}}}
\newcommand{\vonpn}{\ensuremath{v_{\omega N,\pi N}}}
\newcommand{\vpnpn}{\ensuremath{v_{\pi N,\pi N}}}
\newcommand{\vonon}{\ensuremath{v_{\omega N,\omega N}}}
\newcommand{\vrngn}{\ensuremath{v_{\rho N,\gamma N}}}
\newcommand{\tjtmbmb}{\ensuremath{t^{JT}_{M'B',MB}}}
\newcommand{\tjlsmngn}{\ensuremath{t^{JT}_{L'S'M'N',\lag\lN T_{N,z}}}}
\newcommand{\tjlsmbgn}{\ensuremath{t^{JT}_{LSMB,\lag \lN T_{N,z}}}}
\newcommand{\vjlsmngn}{\ensuremath{v^{JT}_{L'S'M'N',\lag \lN T_{N,z}}}}
\newcommand{\vjlsmbgn}{\ensuremath{v^{JT}_{LSMB,\lag \lN T_{N,z}}}}
\newcommand{\tjlsmnmb}{\ensuremath{t^{JT}_{L'S'M'N',LSMB}}}
\newcommand{\Tjlsmbmb}{\ensuremath{T^{JT}_{LSMB,L'S'M'B'}}}
\newcommand{\tjlsmbmb}{\ensuremath{t^{JT}_{LSMB,L'S'M'B'}}}
\newcommand{\tjlsmnpn}{\ensuremath{t^{JT}_{L'S'M'N',\ell \pi N}}}
\newcommand{\tjlsmbpn}{\ensuremath{t^{JT}_{LSMB,\ell \pi N}}}
\newcommand{\vjlsmnpn}{\ensuremath{v^{JT}_{L'S'M'N',\ell \pi N}}}
\newcommand{\vjlsmnmb}{\ensuremath{v^{JT}_{L'S'M'N',LSMB}}}
\newcommand{\vjlsmbpn}{\ensuremath{v^{JT}_{LSMB,\ell \pi N}}}
\newcommand{\Tjlsmngn}{\ensuremath{t^{R,JT}_{L'S'M'N',\lag\lN T_{N,z}}}}
\newcommand{\Tjlsmbgn}{\ensuremath{t^{R,JT}_{LSMB,\lag \lN T_{N,z}}}}
\newcommand{\Tfjlsmbgn}{\ensuremath{T^{JT}_{LSMB,\lag \lN T_{N,z}}}}
\newcommand{\Tjlsmnmb}{\ensuremath{t^{R,JT}_{L'S'M'N',LSMB}}}
\newcommand{\Tjlsmnpn}{\ensuremath{t^{R,JT}_{L'S'M'N',\ell \pi N}}}
\newcommand{\Tjlsmbpn}{\ensuremath{t^{R,JT}_{LSMB,\ell \pi N}}}
\newcommand{\Gbjlsi}{\ensuremath{{\Gamma}^{JT}_{LSMB,N^*_i}}}
\newcommand{\Gbjlspi}{\ensuremath{{\Gamma}^{JT}_{L'S'M'B',N^*_i}}}
\newcommand{\Gjlsi}{\ensuremath{\overline{\Gamma}^{JT}_{LSMB,N^*_i}}}
\newcommand{\Gijls}{\ensuremath{\overline{\Gamma}^{JT}_{N^*_i,LSMB}}}
\newcommand{\Gbijls}{\ensuremath{{\Gamma}^{JT}_{N^*_i,LSMB}}}
\newcommand{\Gjpn}{\ensuremath{\overline{\Gamma}^{JT}_{N^*_j,\ell\pn}}}
\newcommand{\Gign}{\ensuremath{\overline{\Gamma}^{JT}_{N^*_i,\lag\lN T_{N,z}}}}
\newcommand{\Gbign}{\ensuremath{{\Gamma}^{JT}_{N^*_i,\lag\lN T_{N,z}}}}
\newcommand{\Gjlsj}{\ensuremath{\overline{\Gamma}^{JT}_{LSMB,N^*_j}}}
\newcommand{\Gjem}{\ensuremath{\overline{\Gamma}^{JT}_{N^*_j,\lag\lN T_{N,z}}}}
\newcommand{\Ljtlsmbn}{\ensuremath{\Lambda^{JT}_{N^*LSMB}}}
\newcommand{\Drij}{\ensuremath{\mathcal{D}^{-1}_{ij}}}
\newcommand{\Mbres}{\ensuremath{M^{(0)}_{N^*}}}
\newcommand{\Cjtnlsmb}{\ensuremath{C^{JT}_{N^*LSMB}}}
\newcommand{\Ljtnlsmb}{\ensuremath{\Lambda^{JT}_{N^*LSMB}}}
\newcommand{\knstar}{\ensuremath{k_{N^*}}}
\newcommand{\vonen}{\ensuremath{v_{\omega N,\eta N}}}
\newcommand{\vonpd}{\ensuremath{v_{\omega N,\pi\Delta}}}
\newcommand{\vonsn}{\ensuremath{v_{\omega N,\sigma N}}}
\newcommand{\vonrn}{\ensuremath{v_{\omega N,\rho N}}}
\newcommand{\gnon}{\ensuremath{\gamma N\to \omega N}}
\newcommand{\gnpn}{\ensuremath{\gamma N\to \pi N}}
\newcommand{\gnky}{\ensuremath{\gamma N\to KY}}
\newcommand{\enepn}{\ensuremath{e N\to e'\pi N}}
\newcommand{\gnen}{\ensuremath{\gamma N\to \eta N}}
\newcommand{\gpop}{\ensuremath{\gamma p\to \omega p}}
\newcommand{\gpep}{\ensuremath{\gamma p\to \eta p}}
\newcommand{\gpepp}{\ensuremath{\gamma p\to \eta' p}}
\newcommand{\gnten}{\ensuremath{\gamma n\to \eta n}}
\newcommand{\pzp}{\ensuremath{\pi^0 p}}
\newcommand{\ppln}{\ensuremath{\pi^+ n}}
\newcommand{\pmp}{\ensuremath{\pi^- p}}
\newcommand{\pzn}{\ensuremath{\pi^0 n}}
\newcommand{\gppzp}{\ensuremath{\gamma p\to \pi^0 p}}
\newcommand{\gpppn}{\ensuremath{\gamma p\to \pi^+ n}}
\newcommand{\gnpmp}{\ensuremath{\gamma n\to \pi^- p}}
\newcommand{\gnpzn}{\ensuremath{\gamma n\to \pi^0 n}}
\newcommand{\gppzep}{\ensuremath{\gamma p\to \pi^0 \eta p}}
\newcommand{\pnen}{\ensuremath{\pi N\to \eta N}}
\newcommand{\pnon}{\ensuremath{\pi N\to \omega N}}
\newcommand{\pnmb}{\ensuremath{\pi N\to MB}}
\newcommand{\gnmb}{\ensuremath{\gamma N\to M\!B}}
\newcommand{\onon}{\ensuremath{\omega N\to \omega N}}
\newcommand{\pmpon}{\ensuremath{\pi^- p\to \omega n}}
\newcommand{\pnpn}{\ensuremath{\pi N\to \pi N}}
\newcommand{\pnppn}{\ensuremath{\pi N\to\pi\pi N}}
\newcommand{\knkn}{\ensuremath{K N\to K N}}
\newcommand{\nnnn}{\ensuremath{N N\to N N}}
\newcommand{\pDpD}{\ensuremath{\pi D\to \pi D}}
\newcommand{\pDpp}{\ensuremath{\pi^+ D\to pp}}
\newcommand{\Gon}{\ensuremath{G_{0,\omega N}}}
\newcommand{\Gpn}{\ensuremath{G_{0,\pi N}}}
\newcommand{\rhomb}{\ensuremath{\rho_{MB}}}
\newcommand{\rhoon}{\ensuremath{\rho_{\omega N}}}
\newcommand{\rhopn}{\ensuremath{\rho_{\pi N}}}
\newcommand{\kon}{\ensuremath{k_{\omega N}}}
\newcommand{\kpn}{\ensuremath{k_{\pi N}}}
\newcommand{\Gmb}{\ensuremath{G_{0,MB}}}
\newcommand{\Tmbgn}{\ensuremath{T_{MB,\gamma N}}}
\newcommand{\vmbpgn}{\ensuremath{v_{M'B',\gamma N}}}
\newcommand{\pntpn}{\ensuremath{\pi N\!\to\!\pi N}}
\newcommand{\pnten}{\ensuremath{\pi N\!\to\!\eta N}}
\newcommand{\pnton}{\ensuremath{\pi N\!\to\!\omega N}}
\newcommand{\epos}{\ensuremath{\slashed{\epsilon}_{\lambda_\omega}}}
\newcommand{\epo}{\ensuremath{{\epsilon}_{\lambda_\omega}}}
\newcommand{\elevi}{\ensuremath{{\epsilon}_{\alpha\beta\gamma\delta}}}
\newcommand{\eps}{\ensuremath{\epsilon}}
\newcommand{\krho}{\ensuremath{\kappa_\rho}}
\newcommand{\komg}{\ensuremath{\kappa_\omega}}
\newcommand{\komga}{\ensuremath{\kappa^t_\omega}}
\newcommand{\doh}{\ensuremath{d^{(\half)}_{\lambda'\lambda}}}
\newcommand{\dohm}{\ensuremath{d^{(\half)}_{-\lambda,-\lambda'}}}
\newcommand{\dohmo}{\ensuremath{d^{(\half)}_{\lambda',-\half}}}
\newcommand{\dohpo}{\ensuremath{d^{(\half)}_{\lambda',+\half}}}
\newcommand{\Lor}[2]{\ensuremath{\Lambda^{#1}_{\hphantom{#1}{#2}}}}
\newcommand{\ILor}[2]{\ensuremath{\Lambda_{#1}^{\hphantom{#1}{#2}}}}
\newcommand{\LorT}[2]{\ensuremath{[\Lambda^T]^{#1}_{\hphantom{#1}{#2}}}}
\newcommand{\dsdo}{\ensuremath{{\frac{d\sigma}{d\Omega}}}}
\newcommand{\dsdol}{\ensuremath{{\frac{d\sigma}{d\Omega_l}}}}
\newcommand{\dsdolo}{\ensuremath{{\frac{d\sigma}{d\Omega_{l,1}}}}}
\newcommand{\dsdoc}{\ensuremath{{\frac{d\sigma}{d\Omega_c}}}}
\newcommand{\dsdon}{\ensuremath{{{d\sigma}/{d\Omega}}}}
\newcommand{\dspdo}{\ensuremath{{\frac{d\sigma_\pi}{d\Omega}}}}
\newcommand{\dsgdo}{\ensuremath{{\frac{d\sigma_\gamma}{d\Omega}}}}
\newcommand{\chipd}{\ensuremath{\chi^2/N_d}}
\newcommand{\chipda}{\ensuremath{\chi^2(\alpha)/N_d}}
\newcommand{\bpop}{\ensuremath{\bvec{p}'_1}}
\newcommand{\bptp}{\ensuremath{\bvec{p}'_2}}
\newcommand{\bpip}{\ensuremath{\bvec{p}'_i}}
\newcommand{\bpo}{\ensuremath{\bvec{p}_1}}
\newcommand{\bpt}{\ensuremath{\bvec{p}_2}}
\newcommand{\bpi}{\ensuremath{\bvec{p}_i}}
\newcommand{\bqo}{\ensuremath{\bvec{q}_1}}
\newcommand{\bqt}{\ensuremath{\bvec{q}_2}}
\newcommand{\bqi}{\ensuremath{\bvec{q}_i}}
\newcommand{\bQ}{\ensuremath{\bvec{Q}}}
\newcommand{\bq}{\ensuremath{\bvec{q}}}
\newcommand{\ketq}{\ensuremath{\ket{\bqo,\bqt}}}
\newcommand{\ketqc}{\ensuremath{\ket{\bQ,\bq}}}
\newcommand{\bP}{\ensuremath{\bvec{P}}}
\newcommand{\bPp}{\ensuremath{\bvec{P}'}}
\newcommand{\bpr}{\ensuremath{\bvec{p}}}
\newcommand{\bprp}{\ensuremath{\bvec{p}'}}
\newcommand{\ketPsiq}{\ensuremath{\ket{\Psi_{\bq}^{(\pm)}}}}
\newcommand{\ketPsiqQ}{\ensuremath{\ket{\Psi_{\bQ,\bq}^{(\pm)}}}}
\newcommand{\Ld}{\ensuremath{\mathcal{L}}}
\newcommand{\ps}{\mbox{ps}}
\newcommand{\fndp}{f_{N\Delta\pi}}
\newcommand{\fndr}{f_{N\Delta\rho}}
\newcommand{\said}{{\sc said}}
\newcommand{\ret}{\ensuremath{\langle{\tt ret}\rangle}}
\newcommand{\ddf}[1]{\ensuremath{\delta^{(#1)}}}
\newcommand{\Tpp}{\ensuremath{T_{\pi\pi}}}
\newcommand{\Kpp}{\ensuremath{K_{\pi\pi}}}
\newcommand{\Tpe}{\ensuremath{T_{\pi\eta}}}
\newcommand{\Kpe}{\ensuremath{K_{\pi\eta}}}
\newcommand{\Tep}{\ensuremath{T_{\eta\pi}}}
\newcommand{\Kep}{\ensuremath{K_{\eta\pi}}}
\newcommand{\Tee}{\ensuremath{T_{\eta\eta}}}
\newcommand{\Kee}{\ensuremath{K_{\eta\eta}}}
\newcommand{\Tpig}{\ensuremath{T_{\pi\gamma}}}
\newcommand{\Kpig}{\ensuremath{K_{\pi\gamma}}}
\newcommand{\oKpig}{\ensuremath{\overline{K}_{\pi\gamma}}}
\newcommand{\tKpig}{\ensuremath{\tilde{K}_{\pi\gamma}}}
\newcommand{\Teg}{\ensuremath{T_{\eta\gamma}}}
\newcommand{\Keg}{\ensuremath{K_{\eta\gamma}}}
\newcommand{\Kab}{\ensuremath{K_{\alpha\beta}}}
\newcommand{\R}{\ensuremath{\mathbb{R}}}
\newcommand{\C}{\ensuremath{\mathbb{C}}}
\newcommand{\Ezp}{\ensuremath{E^{\pi}_{0+}}}
\newcommand{\Eze}{\ensuremath{E^{\eta}_{0+}}}
\newcommand{\Ga}{\ensuremath{\Gamma_\alpha}}
\newcommand{\Gb}{\ensuremath{\Gamma_\beta}}
\newcommand{\RH}{\ensuremath{\mathcal{R}\!\!-\!\!\mathcal{H}}}
\newcommand{\calT}{\mathcal{T}}
\newcommand{\maid}{{\sc maid}}
\newcommand{\Kbar}{\ensuremath{\overline{K}}}
\newcommand{\zbar}{\ensuremath{\overline{z}}}
\newcommand{\kbar}{\ensuremath{\overline{k}}}
\newcommand{\dom}{\ensuremath{\mathcal{D}}}
\newcommand{\domi}[1]{\ensuremath{\mathcal{D}_{#1}}}
\newcommand{\pbar}{\ensuremath{\overline{p}}}
\newcommand{\Nab}{\ensuremath{N_{\alpha\beta}}}
\newcommand{\Nee}{\ensuremath{N_{\eta\eta}}}
\newcommand{\dth}[1]{\delta^{(3)}(#1)}
\newcommand{\dfo}[1]{\delta^{(4)}(#1)}
\newcommand{\intk}{\int\!\!\frac{d^3\! k}{(2\pi)^3}}
\newcommand{\intkg}{\int\!\!{d^3\! k_\gamma}}
\newcommand{\intks}{\int\!\!{d^3\! k_\sigma}}
\newcommand{\nch}{\ensuremath{N_{\mbox{ch}}}}
\newcommand{\nc}{\ensuremath{N_{ch}}}
\newcommand{\re}{\ensuremath{\mbox{Re }\!}}
\newcommand{\im}{\ensuremath{\mbox{Im }\!}}
\newcommand{\EetaS}{\ensuremath{E^\eta_{0+}}}
\newcommand{\EpiS}{\ensuremath{E^\pi_{0+}}}
\newcommand{\tobull}{\ensuremath{\to}}
\newcommand{\Kcm}{\ensuremath{K_{CM}}}
\newcommand{\lra}{\ensuremath{\leftrightarrow}}

\newcommand{\gn}{\ensuremath{\gamma N}}
\newcommand{\gp}{\ensuremath{\gamma p}}
\newcommand{\geta}{\ensuremath{\gamma \eta}}
\newcommand{\pp}{\ensuremath{pp}}
\newcommand{\pn}{\ensuremath{\pi N}}
\newcommand{\phn}{\ensuremath{\pi d}}
\newcommand{\en}{\ensuremath{\eta N}}
\newcommand{\epn}{\ensuremath{\eta' N}}
\newcommand{\pD}{\ensuremath{\pi \Delta}}
\newcommand{\sn}{\ensuremath{\sigma N}}
\newcommand{\rn}{\ensuremath{\rho N}}
\newcommand{\on}{\ensuremath{\omega N}}
\newcommand{\ppn}{\ensuremath{\pi\pi N}}
\newcommand{\pipi}{\ensuremath{\pi\pi}}
\newcommand{\kn}{\ensuremath{KN}}
\newcommand{\ky}{\ensuremath{KY}}
\newcommand{\kl}{\ensuremath{K\Lambda}}
\newcommand{\ks}{\ensuremath{K\Sigma}}
\newcommand{\bn}{\ensuremath{eN}}
\newcommand{\pR}{\ensuremath{\pi N^*}}
\newcommand{\bpn}{\ensuremath{e\pi N}}
\newcommand{\fpo}{\ensuremath{5\oplus 1}}
\newcommand{\faoe}{{\sc FA08}}
\newcommand{\fpoe}{{\sc FP08}}
\newcommand{\fsoe}{{\sc FS08}}
\newcommand{\psic}{\ensuremath{\psi_{n\kappa jm}}}
\newcommand{\hi}[1]{\ensuremath{H_I(t_{#1})}}
\newcommand{\dcp}{\ensuremath{\mathcal{D}}}
\newcommand{\pptopp}{\ensuremath{\pipi\to\pipi}}
\newcommand{\pntoppn}{\ensuremath{\pn\to\ppn}}
\newcommand{\ovlt}{\ensuremath{\overline{t}}}
\newcommand{\smat}{\ensuremath{e^{-i\int_{-\infty}^\infty\!dt\,H_I(t)}}}
\newcommand{\pder}[2][]{\frac{\partial#1}{\partial#2}}
\newcommand{\der}[2][]{\frac{d#1}{d#2}}

\newcommand{\Rhmat}{$R$-matrix}
\newcommand{\Rmat}{$R$ matrix}
\newcommand{\svb}{\ensuremath{\langle\sigma v\rangle}}
\newcommand{\redG}{\ensuremath{\tilde{G}}}

\newcommand{\mde}{\ensuremath{d}}
\newcommand{\de}{d}
\newcommand{\mtriton}{\ensuremath{t}}
\newcommand{\triton}{t}
\newcommand{\mHiii}{\ensuremath{^3\!H}}
\newcommand{\Hiii}{$^{3}$H}
\newcommand{\mHeiii}{\ensuremath{^3\!H\!e}}
\newcommand{\Heiii}{$^{3}$He}
\newcommand{\mHeiv}{\ensuremath{^4\!H\!e}}
\newcommand{\Heiv}{$^4$He}
\newcommand{\mHev}{\ensuremath{^5\!H\!e}}
\newcommand{\Hev}{\ensuremath{^5}He}
\newcommand{\mHevi}{\ensuremath{^6\!He}}
\newcommand{\Hevi}{\ensuremath{^6}He}
\newcommand{\mLivi}{\ensuremath{^6\!Li}}
\newcommand{\Livi}{\ensuremath{^6}Li}
\newcommand{\mLivii}{\ensuremath{^7\!Li}}
\newcommand{\Livii}{\ensuremath{^7}Li}
\newcommand{\mBevii}{\ensuremath{^7\!Be}}
\newcommand{\Bevii}{\ensuremath{^7}Be}
\newcommand{\mBeviii}{\ensuremath{^8\!Be}}
\newcommand{\Beviii}{\ensuremath{^8}Be}
\newcommand{\mBix}{\ensuremath{^9\!B}}
\newcommand{\Bix}{\ensuremath{^9}B}
\newcommand{\mCxiii}{\ensuremath{^{13}\!C}}
\newcommand{\Cxiii}{\ensuremath{^{13}}C}
\newcommand{\mCxiv}{\ensuremath{^{14}\!C}}
\newcommand{\Cxiv}{\ensuremath{^{14}}C}

\newcommand{\EDA}{{\tt EDA}}
\newcommand{\ttinp}{{\tt input}}
\newcommand{\ttout}{{\tt output}}
\newcommand{\ttdat}{{\tt data}}
\newcommand{\ttpar}{{\tt par}}
\newcommand{\ttnam}{{\tt namelst}}
\newcommand{\tthol}{{\tt hold}}
\newcommand{\tthne}{{\tt hnew}}
\newcommand{\tteda}{{\tt eda}}
\newcommand{\ttsou}{{\tt sout}}
\newcommand{\ttpan}{{\tt parnew}}
\newcommand{\tturd}{{\tt urd}}
\newcommand{\ttfen}{{\tt fengy}}
\newcommand{\ttfet}{{\tt fengy2}}
\newcommand{\ttplo}{{\tt plotf}}
\newcommand{\ttpao}{{\tt par1r}}
\newcommand{\ttfdb}{{\tt fdbdu}}
\newcommand{\unk}{{\bf unk}}
\newcommand{\ttchr}{{\tt changri}}

\newcommand{\itPFP}{\textit{Physics for Future Presidents}}
\newcommand{\itaPFP}{\textit{PFP}}
\newcommand{\prle}{\textit{PR}\textbf{97}}
\newcommand{\lcd}{---\;}

\setcounter{page}{1}

\title{
\qquad \\ \qquad \\ \qquad \\  \qquad \\  \qquad \\ \qquad \\ 
R-matrix analysis of reactions in the $^{9}$B compound system
}

\author{M.\ Paris}
\email[Corresponding author, electronic address:\\ ]{mparis@lanl.gov}
\author{G.\ Hale}
\author{A.\ Hayes-Sterbenz} 
\author{G.\ Jungman} 
\affiliation{T-2 Theoretical Division, Los Alamos Nuclear Laboratory, MS B283, Los Alamos, New Mexico 87545, USA}

\date{\today} 

\begin{abstract}
{
Recent activity in solving the `lithium problem' in big bang
nucleosynthesis has focused on the role that putative resonances may
play in resonance-enhanced destruction of \Livii. Particular attention
has been paid to the reactions involving the \Bix\ compound nuclear
system, \de+\Bevii$\to$\Bix.  These reactions
are analyzed via the multichannel, two-body unitary $R$-matrix method
using code ({\it EDA}) developed by Hale and collaborators. We employ
much of
the known elastic and reaction data, in a four-channel treatment. The
data include elastic \Heiii+\Livi\ differential cross sections from
0.7 to 2.0 MeV, integrated reaction cross sections for energies from
0.7 to 5.0 MeV for \Livi(\Heiii,p)\Beviii$^*$ and from 0.4 to 5.0 MeV
for the \Livi(\Heiii,d)\Bevii\ reaction. Capture data have been added
to an earlier analysis with integrated cross section measurements
from 0.7 to 0.825 MeV for \Livi(\Heiii,$\gamma$)\Bix. The resulting
resonance parameters are compared with tabulated values, and
previously unidentified resonances are noted. Our results show that
there are no near \de+\Bevii\ threshold resonances with widths that
are 10's of keV and reduce the likelihood that a resonance-enhanced
mass-7 destruction mechanism, as suggested in recently published work,
can explain the \Livii\ problem.
}
\end{abstract}
\maketitle


\lhead{$R$-matrix analysis\ldots}
\chead{NUCLEAR DATA SHEETS}
\rhead{M.\ Paris \textit{et al.}}
\lfoot{}
\rfoot{}
\renewcommand{\footrulewidth}{0.4pt}

\section{ INTRODUCTION}
\label{sec:intro}
Calculations of the abundance of \Livii\cite{Fields:2011zzb}
overestimate the value extracted from observations of low-metallicity
halo dwarf stars\cite{Spite:1982dd}, where the stellar dynamics are
supposed to be sufficiently understood to isolate the primordial
\Livii\ component. The discrepancy with this (and
another\cite{Gonz:2009gl}) observation by a factor of $2.2\to 4.2$
corresponds to a deviation of $4.5\sigma \to 5.5\sigma$, a result that has
only become more severe with time.  It is essential to determine
the nature of this discrepancy as big-bang nucleosynthesis (BBN)
probes conditions of the very early universe and our understanding of
physical laws relevant in an extreme environment.

Recent attention has focused on the role of reactions that destroy
$A=7$ nuclei at early times $\lesssim 1$ s in the big-bang
environment\cite{Cyburt:2009cf,Chakraborty:2010zj}. 
The authors of Ref.\cite{Cyburt:2009cf}, citing the
TUNL-Nuclear Data Group (NDG) evaluation tables\cite{Tilley:2004zz}, 
(See Table \ref{tab:tunl}.) 
conjecture that the
putative $\sfhalf^+$ resonance near $16.7$ MeV may enhance the
destruction of \Bevii\ through reactions like
\Bevii(d,p)$\alpha\alpha$ and \Bevii(d,$\gamma$)\Bix\ if the resonance
parameters are within given ranges. These studies employ the Wigner
limit\cite{Teichmann:1952aa} to determine an upper bound on the
contribution of resonances, particularly \Bix, to a resonant
enhancement in reactions that destroy mass-7 nuclides, \Livii\ in
particular. Because there is a paucity of data in the region
near the d+\Bevii\ threshold where the assumed $\sfhalf^+$ \Bix\ 
resonance inhabits, we wondered if the existing data may indicate the
presence of such a resonance if a multichannel, unitary $R$-matrix
evaluation is pursued.

\begin{table}
   \begin{tabular}{|c|c|c|c|}
      \hline
      $E_x(\mbox{MeV}\pm\mbox{keV})$ &
			  $J^\pi; T$ &
      $\Gamma_{\mbox{cm}}$(keV)&
      Decay\\
      \hline
      $16.024 \pm 25$ &
      $T=\left(\shalf\right)$ &
		 $180 \pm 16$ &
		\\
		$16.71 \pm 100$ &
	  $(\sfhalf^+);(\shalf)$&
				       &
		\\
		$17.076\pm 4$&
		$\shalf^-;\sthalf$&
			 $22\pm 5$&
		($\gamma$,\Heiii)\\
		$17.190\pm 25$&
			      &
	    $120\pm 40$&
		p, d, \Heiii\\
		$17.54\pm 100$ &
       ($\sshalf^+$);($\shalf$)&
					     &
		\\
		$17.637\pm 10$&
			      &
	      $71\pm 8$&
		p, d, \Heiii, $\alpha$\\
		\hline
   \end{tabular}
   \caption{\label{tab:tunl}The TUNL-NDG/ENSDF resonances in
	 the \Bix\ compound nuclear system\cite{Tilley:2004zz}
         for resonances
	 that are low-lying with respect to the d-\Bevii\ threshold,
      which occurs at 16.4901 MeV.}
\end{table}

Our motivation for the present study of the \Bix\ compound system is
two-fold. The continuing light nuclear reaction program at Los Alamos
National Laboratory, T-2 Theoretical Division provides light nuclear
data for an array of end users, including the ENDF and ENSDF
communities. Moreover, we are interested in updating the evaluation of
the \Bix\ compound system to address the key question outlined above
for BBN: does a resonance near the d+\Bevii\ threshold cause an
increase in the destruction of mass-7 nuclides in the early universe
and possibly explain the \Livii\ overprediciton problem?

\section{The $R$-matrix formalism and EDA code}
\label{sec:Rmat}
The $R$-matrix approach\cite{Kapur:1938kp,Wigner:1946zz,Wigner:1947zz}
is a unitary, multichannel
parametrization that has proven useful for an array of nuclear
reaction phenomenology, particularly for light nuclei\cite{Hale:2008nd}. 
We give only a brief description here and refer to the literature for
a more complete description\cite{Lane:1948zh,Lane:1966zz}.

We consider only $2\to 2$ body scattering and reaction processes for
light nuclear systems. Configuration space is partitioned into an
interior, strongly interacting region and an exterior, Coulomb or
non-polarizing interaction region by giving a channel radius, $a_c$
for each two-body channel. The boundary of separation of these
regions is the channel surface, $\mathcal{S} = \sum_c \mathcal{S}_c$.

The $R$ matrix is computed as the projection on channel surface
functions, $|c)$ of the Green's function, $G_B =
(H+\mathcal{L}_B-E)^{-1}$
\begin{align}
   R_{c'c} &= (c'|H+\mathcal{L}_B-E)^{-1}|c)
		  = \sum_\lambda \frac{(c'|\lambda)(\lambda|c)}
   {E_\lambda - E},
\end{align}
where $\mathcal{L}_B$ is the Bloch operator, which accounts for the
presence of a boundary condition, $B$ on the channel surface. The
Bloch operator ensures that the operator $H+\mathcal{L}_B$ is a
compact, Hermitian operator having a real, discrete spectrum. The
$R$-matrix parameters, $E_\lambda$ and $\gamma_{\lambda c}=(c\,|\lambda)$ 
describe the spectrum and residues of the resolvent operator; they
are treated as parameters adjusted to fit the observed data. Both
hadronic and electromagnetic ({\em ie.} $\gamma$+\Bix) channels can
be handled in this approach. The transition matrix, $\mathbf{T}$ 
the square of 
which gives the observables (cross section, etc.) of the theory, is
given as
\begin{align}
   \mathbf{T} &= \rho^{1/2}\mathbf{O}^{-1}\mathbf{R}_L\mathbf{O}^{-1}
   \rho^{1/2}-\mathbf{F}\mathbf{O}^{-1},
\end{align}
where $\mathbf{R}_L = (\mathbf{R}^{-1}-\mathbf{L}+\mathbf{B})^{-1}$,
$\mathbf{L}=\rho\mathbf{O}'\mathbf{O}^{-1}$, and
$\mathbf{F}=\mbox{Im }\mathbf{O}$, where $\mathbf{O}$ is the diagonal
matrix of outgoing (Coulomb) wave functions in the exterior region.

The $R$-matrix approach is implemented in the EDA (Energy Dependent
Analysis) code developed by Hale and collaborators\cite{Hale:2008nd}. The
available two-body scattering and reaction data is described by
minimization of the $\chi^2$ function with respect to variation of the
$R$-matrix parameters $E_\lambda$ and $\gamma_{\lambda c}$.

\section{Analysis and results}
\label{sec:ar}
The $R$-matrix configuration, constructed for input into the EDA code,
is given in terms of the included channel partitions (pairs), the $LS$ 
terms for each partition, and the channel radii and boundary
conditions, $B_c$ for each channel.

We have included in the analysis the hadronic channels:
d+\Bevii\ partition with threshold
of 16.5 MeV with up to $D$-waves, \Heiii+\Livi\ at 16.6 MeV up to
$P$-waves, and p+\Beviii$^*$ at 16.7 MeV up to $P$-waves. The channel
radii were constrained to lie in the range between 5.5 fm and 7.5 fm
for these. The electromagnetic $\gamma$+\Bix\ channels included were
$E_1^{3/2}$, $M_1^{5/2}$, $M_1^{3/2}$, $M_1^{1/2}$, $E_1^{5/2}$, and
$E_1^{1/2}$ with a channel radius of 50.0 fm.

The \Bix\ analysis is based upon data gathered from the literature and
stored in the EXFOR/CSISRS database\cite{X4}. We include elastic
differential cross section data for the \Livi+\Heiii\ channel given in
the range of \Heiii\ lab energy, 1.30 MeV$<\!\!E$(\Heiii)$<$1.97 
MeV\cite{Buz79}; integrated cross section data for
\Livi(\Heiii,p)\Beviii$^*$\cite{Elwyn80} where the final state channel
is an average of the excited-states of the quasi-two-body final state
of p+\Beviii$^*$ given, in the range 0.66 MeV $<E$(\Heiii)$< 5.00$ MeV;
integrated cross section for the
\Livi(\Heiii,d)\Bevii\cite{BarrGil65} in the range
0.42 MeV $<E$(\Heiii)$< 4.94$ MeV; and capture data from the 
\Livi+\Heiii\ initial state in the energy range 
0.7 MeV$<\!\!E$(\Heiii)$<0.825$ MeV\cite{Alek78}. 

\begin{figure}
\includegraphics[width=3.5in]{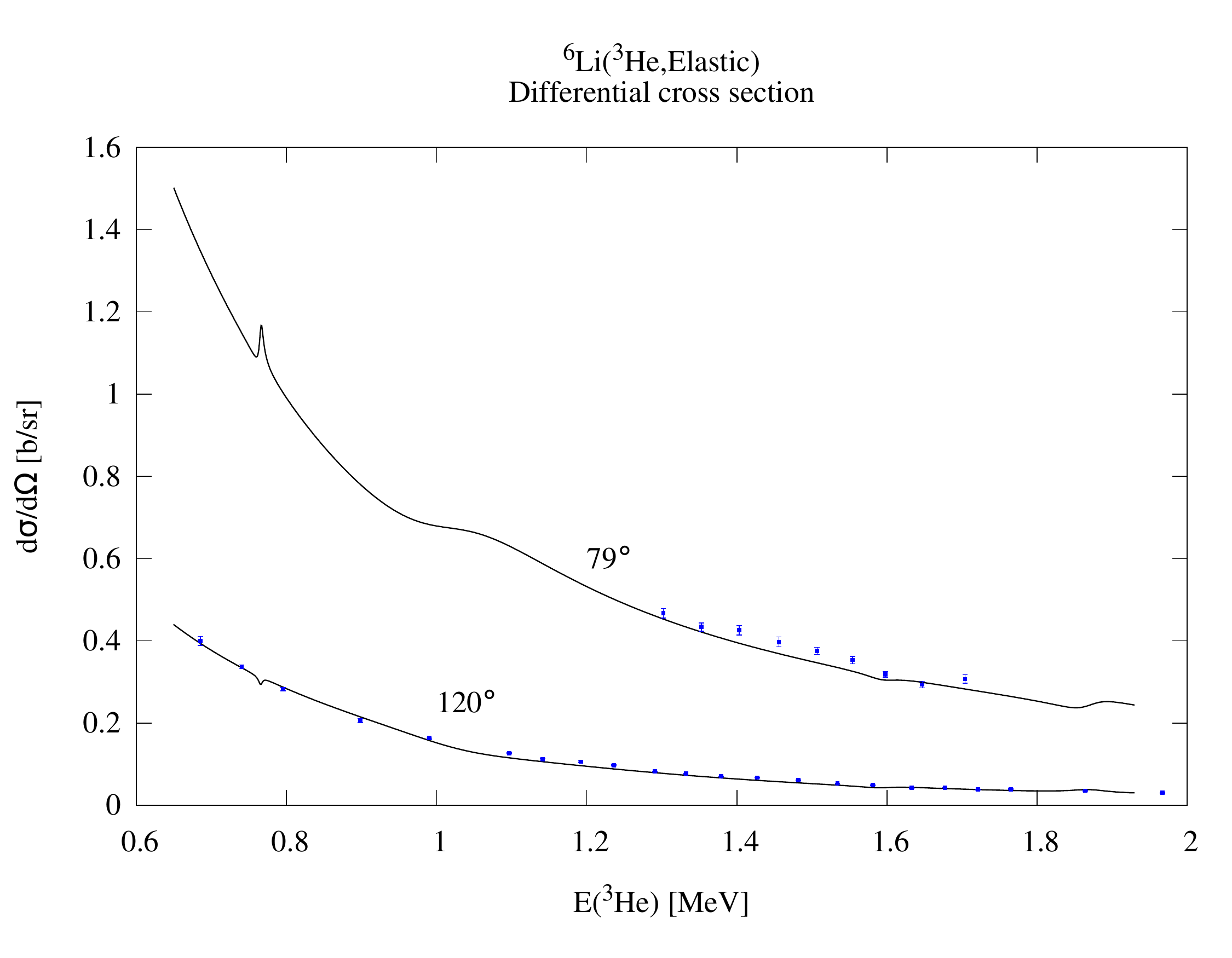}
\caption{\label{fig:el}Elastic scattering data from \cite{Buz79}
plotted against the $R$-matrix fit (solid curve) for center-of-mass
differential cross section vs.\ \Heiii\ lab energy.}
\end{figure}

\begin{figure}
\includegraphics[width=3.5in]{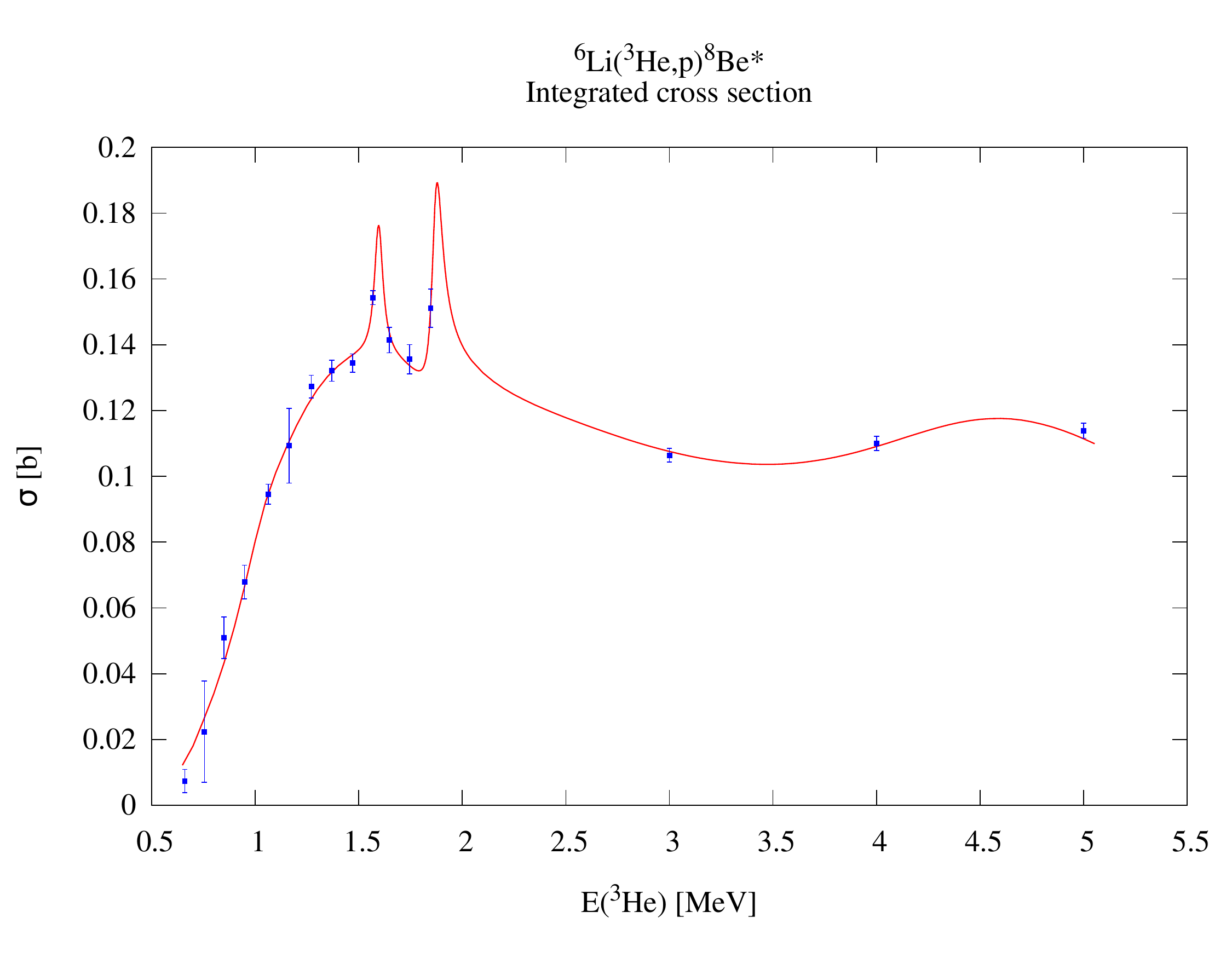}
\caption{\label{fig:8be}Reaction data from \cite{Elwyn80},
integrated cross section vs.\ \Heiii\ lab energy.}
\end{figure}

\begin{figure}
\includegraphics[width=3.5in]{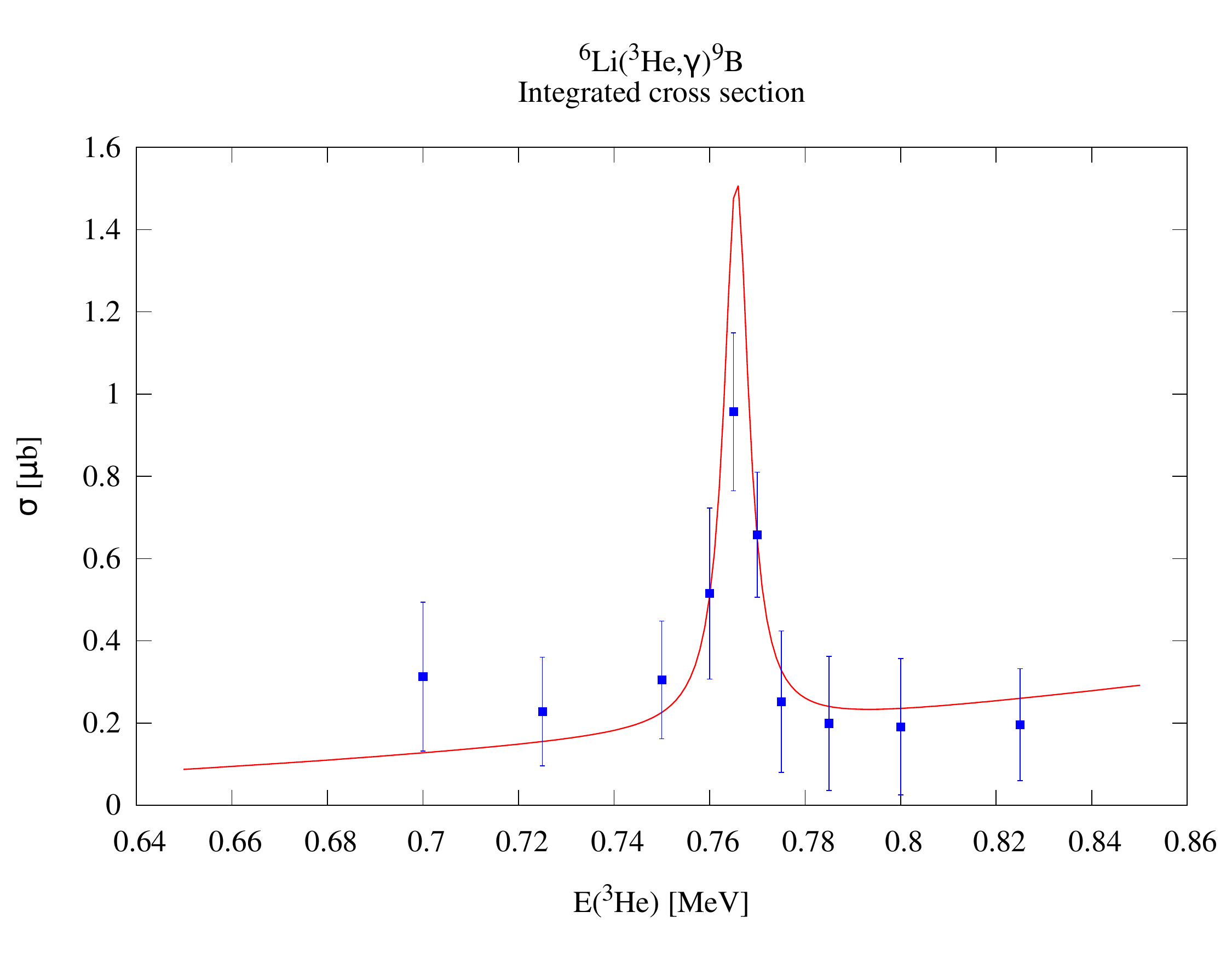}
\caption{\label{fig:cap}Capture data from \cite{Alek78},
integrated cross section vs.\ \Heiii\ lab energy.}
\end{figure}

\begin{table*}[!htb]
\newcolumntype{d}[1]{D{.}{.}{#1}}
\begin{tabular*}{0.8\textwidth}{@{\extracolsep{\fill}}d{4}cd{2}d{4}d{2}d{4}d{2}l}
   \multicolumn{1}{c}{$E_x$(MeV)}&   \multicolumn{1}{c}{$J^\pi$}& \multicolumn{1}{c}{$\Gamma$(keV)}&\multicolumn{1}{c}{Re$E_0$(MeV)}& \multicolumn{1}{c}{Im$E_0$(keV)}&  \multicolumn{1}{c}{$E$(\Heiii)(MeV)}&  \multicolumn{1}{l}{Strength} & \\
 \hline
 16.4754 &  1/2$^-$&   768.46 &  -.1369 &  -384.2   & -0.2054 & 0.06 & weak\\
 17.1132 &  1/2$^-$&     0.14 &  0.5109 &  -0.07    &  0.7664 & 1.00 & strong\\
 17.2012 &  5/2$^-$&   871.63 &  0.5989 &  -435.8   &  0.8984 & 0.40 & weak\\
 17.2809 &  3/2$^-$&   147.78 &  0.6785 &  -73.9    &  1.0178 & 0.77 & strong\\
 17.6754 &  5/2$^+$&    33.33 &  1.0631 &  -16.7    &  1.5947 & 0.98 & strong\\
 17.8462 &  7/2$^+$&  2036.21 &  1.2339 &  -1018.1  &  1.8509 & 0.15 & weak\\
 17.8577 &  3/2$^-$&    42.52 &  1.2454 &  -21.3    &  1.8681 & 0.97 & strong\\
 18.0582 &  3/2$^+$&   767.11 &  1.4459 &  -383.6   &  2.1689 & 0.54 & weak\\
 18.4229 &  1/2$^+$&  5446.32 &  1.8206 &  -2723.2  &  2.7309 & 0.03 & weak\\
 18.6872 &  1/2$^-$& 10278.41 &  2.0749 &  -5139.2  &  3.1124 & 0.15 & weak\\
 19.6192 &  3/2$^-$&  1478.22 &  3.0069 &  -739.1   &  4.5104 & 0.52 & weak\\
 \hline
\end{tabular*}
\caption{\label{tab:res}The resonance structure determined in the
   present 4-channel fit to data as described in the text. The table
   displays the pole location along with
   $J^\pi$ and pole-strength information, as described in the text.}
\end{table*}
Using about 40 parameters, the results of the $\chi^2$ minimization
result in a $T$ matrix which gives the solid curves appearing in
Figs.\ref{fig:el}--\ref{fig:cap},
plotted along with the data obtained from references cited in the
paragraphs above. The fit quality is fair with $\chi^2/\mbox{datum}$
of 1.91, 0.55, 2.38, and 0.37 for Figs.\ref{fig:el}--\ref{fig:cap}, 
respectively. The fit to the capture data, Fig.\ref{fig:cap} has
been folded with a Gaussian acceptance function whose width is 5 keV
to match the quoted energy resolution in Ref.\cite{Alek78}.

The present $R$-matrix parametrization gives a resonance structure
as presented in Table \ref{tab:res}. The resonance poles of the $T$
matrix are determined by diagonalization of the complex
``energy-level'' matrix
\begin{align}
   \label{eqn:elm}
   \mathcal{E}_{\lambda'\lambda}
   &= E_\lambda\delta_{\lambda'\lambda}
   -\sum_c \gamma_{c\lambda'}[L_c(E)-B_c]\gamma_{c\lambda},
\end{align}
where $L_c(E)=r_c(\pd\mathcal{O}/\pd
r_c)\mathcal{O}^{-1}\big|_{r_c=a_c}$, $\mathcal{O}$ is the outgoing
Coulomb wave function, and $B_c$ is the boundary condition given at the
channel radius, $a_c$. Details are given in Ref.\cite{Hale:1987zz}.

The first column of Table \ref{tab:res} gives the real part of the
pole position, $E_0 = E_r - i\Gamma/2$, where $E_0$ is one of the
eigenvalues of the energy-level matrix, Eq.\eqref{eqn:elm} relative
the ground state of \Bix. The spin-parity is given in the second
column. The width $\Gamma$ is the center-of-mass width in keV. The
following column restates the real part of the resonance pole position 
relative the \Heiii+\Livi\ threshold in the center-of-mass. The column 
labeled $E$(\Heiii) is the corresponding lab energy. The `Strength'
function is the ratio of the sum of the channel widths (defined in
Ref.\cite{Hale:1987zz}) divided by the total width,
$\Gamma^{-1}\sum_c\Gamma_c$. Resonances labeled `strong' are clearly
seen in at least one of the figures.

The resonance structure shown in Table \ref{tab:res} differs
significantly from that in Table \ref{tab:tunl}. Possible reasons for
the discrepancy include the fact that the current analysis is the
first, to our knowledge, that includes much of the available data in
the region below $E$(\Heiii)$<3.0$ MeV in a two-body unitary analysis.
Several deductions about the resonance structure in the TUNL/ENSDF
tables rely on associated production of \Bix\
experiments and single-level $R$-matrix
parametrizations\cite{Tilley:2004zz}. While more data, particularly
polarization observables, would constrain the current fit with greater
confidence, the present analysis appears to be the most comprehensive
available that accounts for the available data in a two-body unitary
way.

\begin{figure}[!htb]
\includegraphics[width=3.5in]{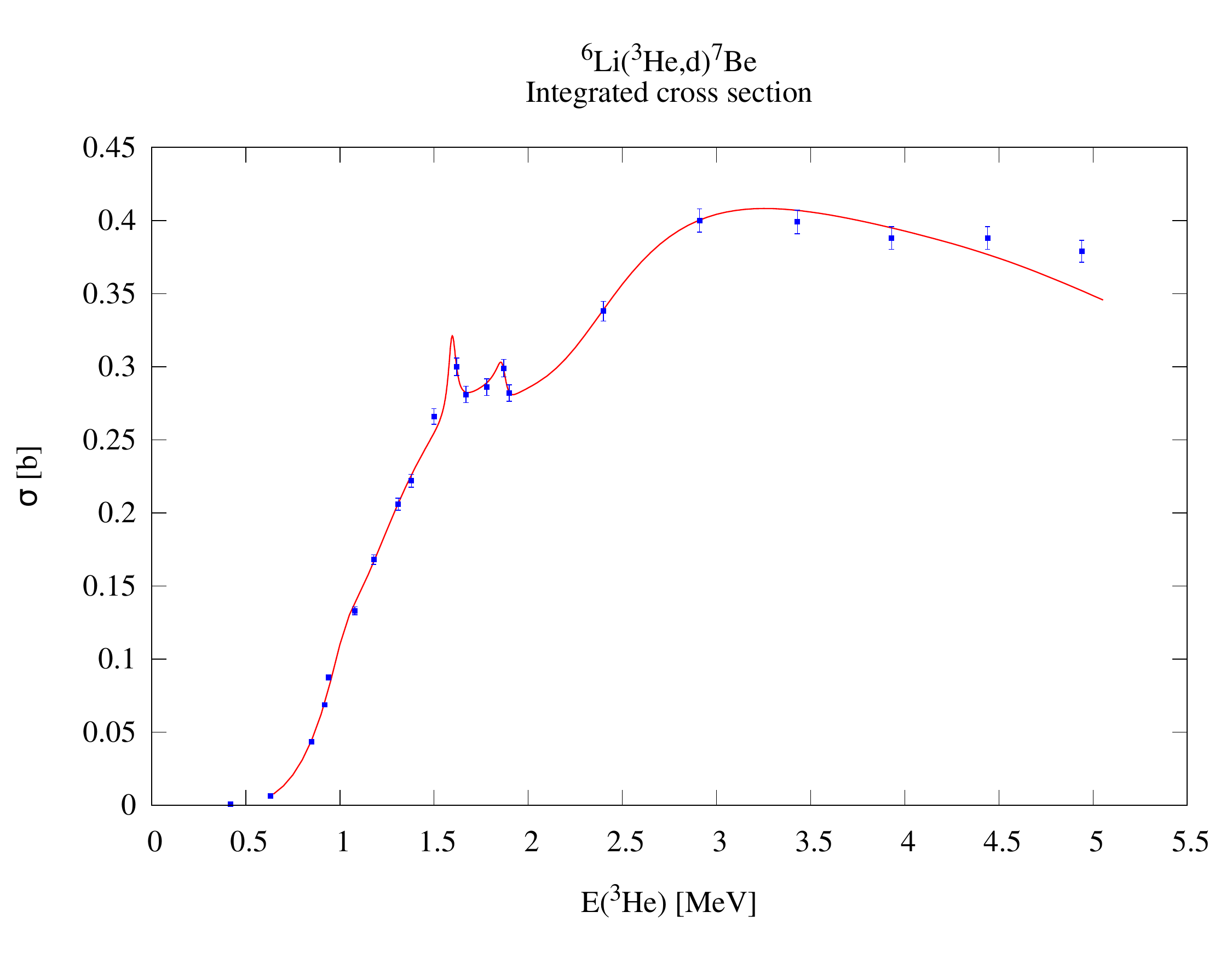}
\caption{\label{fig:d}Reaction data from \cite{BarrGil65},
integrated cross section vs.\ \Heiii\ lab energy.}
\end{figure}

Returning the problem of the overprediction of \Livii\ in current
treatments of BBN, we see that the requirements for a near-threshold
resonance of Refs.\cite{Cyburt:2009cf}
and \cite{Chakraborty:2010zj} 
are difficult to arrange given the resonance structure
of Table \ref{tab:res}. Both of these works require a narrow
resonance, a few 10's of keV in width with a ~100 keV of the
\Heiii+\Livi\ (that is, ~200 keV within the \de+\Bevii) threshold in
order to explain the overproduction of \Livii\ in BBN reaction network
codes\cite{Fields:2011zzb}.

The current study does not conclusively eliminate the possibility of
the mechanism of resonant enhancement of mass-7 destruction. The \Bix\
compound system was identified originally by Cyburt and
Pospelov\cite{Cyburt:2009cf} 
as playing a potential role in the destruction of
\Bevii\ precisely because there isn't much data in the region near the
\de+\Bevii threshold. Our analysis is performed on essentially the
same data that the existent TUNL-NDG analyses\cite{Tilley:2004zz} a were
performed, with the smallest energy probed about 400--500 keV above
the \Heiii+\Livi\ threshold. It might, therefore, be suspected that 
the present data set would give no indication of such a low-lying
resonance. Our experience with $R$ matrix analysis indicates, however,
that a resonance of 10's of keV in width would likely -- but not
certainly -- have contributions `in the tail' to the observables
considered in the present study, particularly in the
\Livi(\Heiii,\de)\Bevii\ integrated cross section of Fig.\ref{fig:d}.

\section{Summary, findings and future work}
\label{sec:summ}
We've studied a possible resonant enhancement of the destruction of
mass-7 (\Bevii, in particular) in BBN scenarios. The near threshold,
narrow state anticipated in Refs.\cite{Cyburt:2009cf} 
and \cite{Chakraborty:2010zj} appear not
to be supported by our multichannel, two-body unitary $R$-matrix
analysis. We have reviewed the $R$-matrix method implemented in the
Los Alamos reaction code for light nuclei, EDA and have
discussed the included data from four channels: elastic \Heiii+\Livi, 
\Livi(\Heiii,p)\Beviii$^*$, \Livi(\Heiii,d)\Bevii\,
and \Livi(\Heiii,$\gamma$)\Bix.

Our analysis determines a resonance structure significantly different
from that published in the 
TUNL-NDG/ENSDF compilation\cite{Tilley:2004zz}, as
can be seen by comparing the results from the present analysis in
Table \ref{tab:res} with the table, Table \ref{tab:tunl} for the
TUNL-NDG/ENSDF analysis. Our immediate objective is to incorporate the
\Beviii$^*$ final states for each excited state (rather than averaging
their contribution as we have done in the present analysis). This will
allow the extension of the present analysis to higher energies and the
incorporation of polarization data\cite{Si66,Si71b}
that we have neglected.

Our findings for the role of a putative resonance in \Bix\ near the
\de+\Bevii\ threshold as envisioned in Refs.\cite{Cyburt:2009cf} 
and \cite{Chakraborty:2010zj}
is that their particular mechanism of resonant enhancement of mass-7
destruction is an unlikely explanation to the \Livii\ problem in BBN,
though low-energy data would allow a more conclusive statement of this
finding or its converse.

This work was carried out under the auspices of the National Nuclear
Security Administration.

\end{document}